\documentclass[english,aip,apl,reprint]{revtex4-1}
\usepackage{units}
\usepackage{color}
\usepackage{amstext}
\usepackage{amssymb}
\usepackage{amsmath}
\usepackage{graphicx}

\begin{document}

\title{Measurement of valley splitting in high-symmetry Si/SiGe quantum dots}

\author{M.~G.~Borselli}
\email{mborselli@hrl.com}
\author{R.~S.~Ross}
\author{A.~A.~Kiselev}
\author{E.~T.~Croke}
\author{K.~S.~Holabird} 
\author{P.~W.~Deelman} 
\author{L.~D.~Warren} 
\author{I.~Alvarado-Rodriguez}  
\author{I.~Milosavljevic} 
\author{F.~C.~Ku} 
\author{W.~S.~Wong} 
\author{A.~E.~Schmitz} 
\author{M.~Sokolich} 
\author{M.~F.~Gyure}
\author{A.~T.~Hunter}

\affiliation{HRL Laboratories,\,LLC, Malibu, CA 90265, USA}


\begin{abstract}
We have demonstrated few-electron quantum dots in Si/SiGe and InGaAs, with occupation number controllable from $N\!=\!0$. These display a high degree of spatial symmetry and identifiable shell structure. Magnetospectroscopy measurements show that two Si-based devices possess a singlet $N\!=\!2$ ground state at low magnetic field, and therefore, the two-fold valley degeneracy is lifted. The valley splittings in these two devices were $270$ and $120~\mu\text{eV}$, suggesting the presence of atomically sharp interfaces in our heterostructures.
\end{abstract}

\maketitle

Quantum dots fabricated in silicon have been a subject of intense interest recently due to the possibility of their use for semiconductor-based quantum information processing.\cite{Eriksson} An important requirement in silicon is the ability to lift the conduction band valley degeneracy,\cite{Eriksson} which is commonly two-fold in the strained Si heterostructure. A lower bound on the valley splitting in a Si quantum dot can be determined by ground state magnetospectroscopy on the first few electron states.\cite{Hada} Recently, this technique has been applied to metal-oxide-semiconductor-based quantum dots\cite{Xiao, Lim, Lim2} and a range of values has been obtained. However, no valley splitting measurements in Si/SiGe quantum dots have been reported, only estimates from transport data in quantum point contacts (QPC).\cite{Goswami,*McGuire} In the Si/SiGe system, electrons are confined against a buried heterointerface as opposed to the $\text{Si/SiO}_2$ interface. This isolates the electrons from surface traps, but leads to potentially different valley physics. In this Letter we report electron addition spectra for accumulation mode Si/SiGe quantum dots, starting from the first electron. Spectra for equivalent devices fabricated in InGaAs provides validation of our numerical simulations and the high axial symmetry provided by this design.  We additionally report measured spin fillings via magnetospectroscopy for two Si/SiGe dots in which clear changes in total spin with magnetic field allow us to extract valley splittings up to $270~\mu\text{eV}$.

\begin{figure}
\begin{centering}
\includegraphics[width=3.2in]{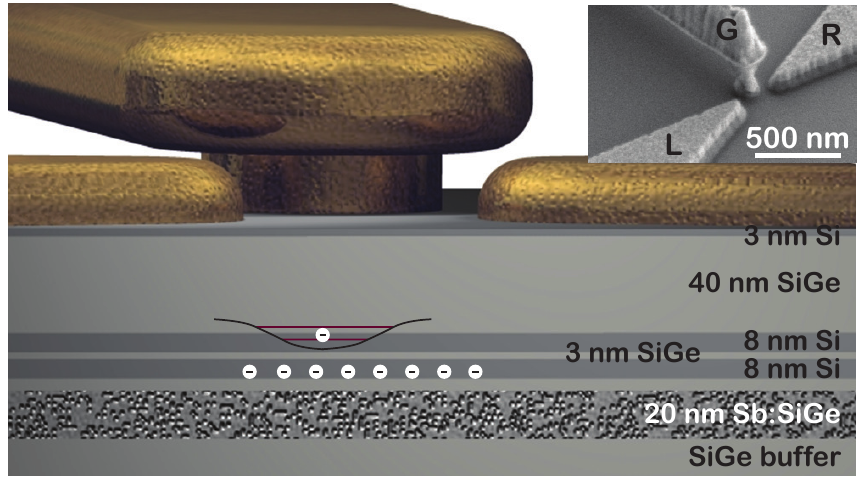}
\par\end{centering}
\caption{\label{fig:Fig1} Schematic representation of the double quantum well device design showing the bridged-gate structure and creation of an accumulation mode dot. (Inset) Scanning electron microscope image of device Si1 showing the air-bridged lead to the circular dot gate (G) and the gates (L,R), which define a QPC in the lower quantum well.}
\end{figure}

The device design is based on a gated double quantum well heterostructure as shown in Fig.~\ref{fig:Fig1}. This design has been realized in both InGaAs, as previously reported by our group,\cite{Croke} and now in Si/SiGe. The Si/SiGe heterostructures were grown on strain-relaxed $\text{Si}_{1-x}\text{Ge}_x$ buffers ($0.26<x<0.34$) and formed two-dimensional electron gases supporting mobilities greater than $20\,000~\text{cm}^2/\text{Vs}$ at $\sim5\times10^{11}~\text{cm}^{-2}$ carrier concentrations. A quantum dot is created in a potential depression in the upper quantum well under the influence of a nominally $100~\text{nm}$ diameter circular forward biased gate that is contacted via an air-bridged (dielectric-bridged for device Si4) lead. A pair of reverse-biased gates form a QPC in the lower quantum well electron gas, within $\sim100~\text{nm}$ of the quantum dot, for charge detection. 

\begin{figure}
\begin{centering}
\includegraphics[width=3.2in]{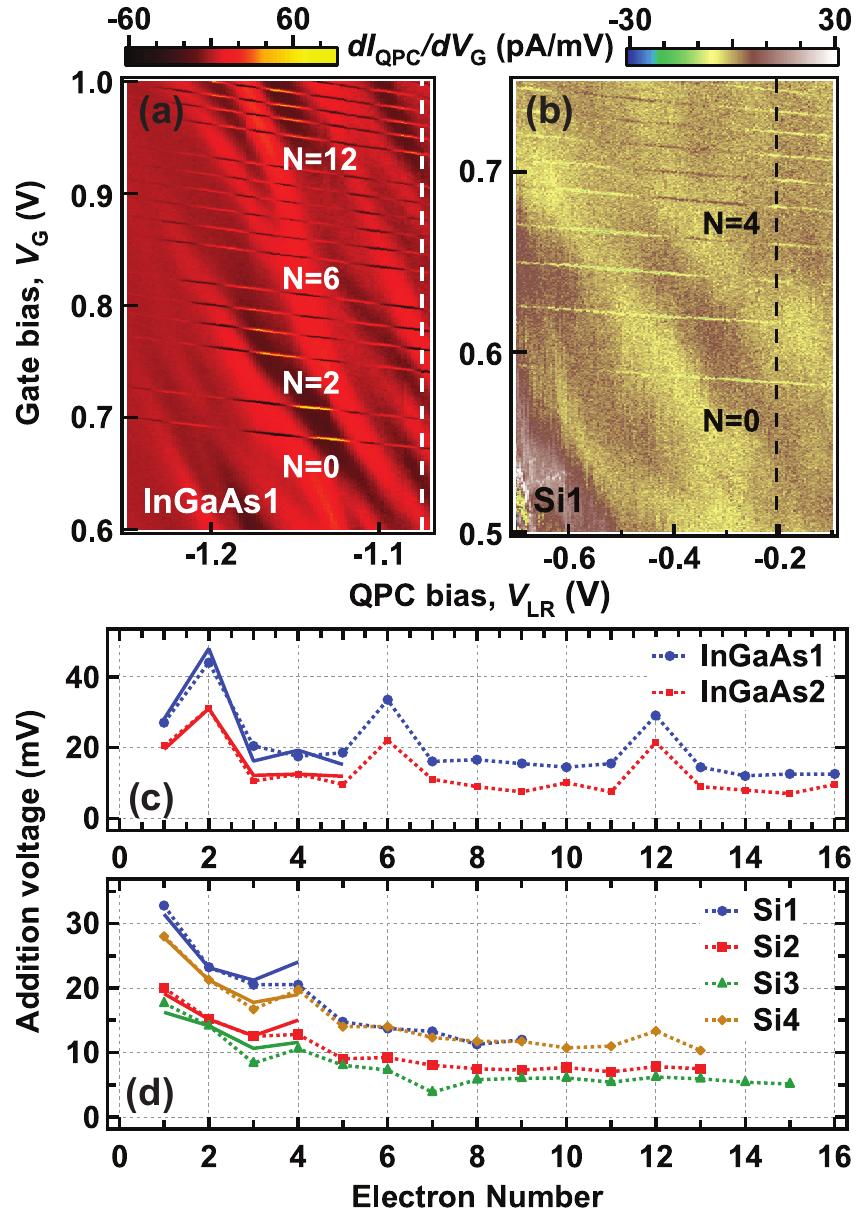}
\par\end{centering}
\caption{\label{fig:Fig2}[(a) and (b)] Differential transconductance ($dI_{QPC}/dV_{G}$) as a function of gate and QPC biases for devices InGaAs1 (left) and Si1 (right). Dashed lines indicate the QPC bias used to extract the addition voltages. [(c) and (d)] Addition voltages for two InGaAs (c) and four Si/SiGe (d) accumulation mode quantum dots. Solid curves are theoretically obtained addition voltages for the first several gaps computed via the FCI method.}
\end{figure}

Six individual devices, having slight variations in dot gate diameter 
and dot-QPC distance, were selected from four Si/SiGe and two InGaAs
wafers and measured in either a $^{3}$He system at an effective electron temperature
of $\sim\!350~\text{mK}$ or a dilution refrigerator at $\sim\!100~\text{mK}$ (device Si4 only). 
Addition spectra were taken using lock-in techniques similar to that described in Croke
\textit{et al}.\cite{Croke} Loading of electrons into the dot was recorded
by sweeping the quantum dot gate bias $V_\text{G}$ and recording
the resulting change in conductance on the constricted current flowing
through the nearby QPC. The dc source-drain
bias was kept to a modest $100~\mu\text{V}$ across the QPC, as larger
currents flowing through the charge sensor were found to broaden the
dot transitions. A differential transconductance $dI_\text{QPC}/dV_\text{G}$
was extracted by measuring the in-phase response of the QPC current
to small sinusoidal or square wave modulation of $V_\text{G}$
(at modulation frequencies well below the tunneling rates for each
respective device). The modulation amplitude was kept as large as
possible without broadening the transitions (typically 0.2 mV$_{\text{rms}}$).
Figures~\ref{fig:Fig2}(a) and \ref{fig:Fig2}(b) show representative high resolution scans
of $dI_\text{QPC}/dV_\text{G}$ in $(V_\text{G},V_\text{LR})$
space of gate and QPC biases, obtained for the InGaAs (left) and Si/SiGe
(right) accumulation mode devices. 

The identification of the last transition as $0\leftrightarrow1$,
in terms of the absolute occupation number $N$, was established by observing
that the tunneling times between the dot and electron reservoir (lower quantum well),
$\sim\!10~\mu\text{s}$, show no significant variation with electron number,
as found by analysis of the time-averaged charge dynamics
for the first few observable transitions. Furthermore, our numerical simulations predict the observed addition spectra including in particular first electron filling voltages and spin fillings.

In Figs.~\ref{fig:Fig2}(c) and \ref{fig:Fig2}(d) the addition voltages are plotted for all 
six measured dots. These data are taken at values
of the QPC biases that provided the largest sustained sensitivity
that captured all of the observed electron loading transitions [in
particular, along dashed vertical lines in Figs.~\ref{fig:Fig2}(a) and \ref{fig:Fig2}(b)].
Addition energies can be obtained from these addition voltages via an independent
measure of the gate lever arm $\alpha$, in which the linewidth of the 
transitions is monitored as a function of sample temperature.\cite{Meirav}
For the series of devices under study $\alpha~\text{[eV/V]}$ is measured to be $0.18$ (InGaAs1), $0.24$ (InGaAs2), $0.14$ (Si1), $0.19$ (Si2), $0.22$ (Si3), within $10\%$ of our numerically simulated values. The simulated value for Si4 is $\alpha = 0.17~\text{eV/V}$.

The solid curves in Figs.~\ref{fig:Fig2}(c) and \ref{fig:Fig2}(d) are simulated addition voltages computed in the framework of the full configuration interaction (FCI) method.\cite{Szabo} They are found as the difference of chemical potentials, $\mu_{N}$, for electron addition and converted to voltage differences via concurrently simulated values of $\alpha$: $V_\text{G}^{N\leftrightarrow(N+1)}-V_\text{G}^{(N-1)\leftrightarrow N}=\left(\mu_{N+1}-\mu_{N}\right)/\alpha$. The chemical potentials, $\mu_{N}\equiv E_\text{Tot}^{N}-E_\text{Tot}^{N-1}$, are computed as the differences between ground state energies of the $N$ and $N-1$ electron systems. Two valleys are explicitly accounted for in the case of Si/SiGe. Accurate simulations using the computationally intensive FCI technique are only available up to five (six) electrons for Si (InGaAs) resulting in the first four (five) addition energies.

First shell filling is observed at $N\!=\!4$ for all four Si dots, as evidenced by the larger addition energy relative to the monotonically decaying Coulomb background. This first shell is much less prominent in Si due to the smaller orbital energy by a factor of $\sqrt{m_\textrm{Si}/m_\textrm{InGaAs}}\sim2.2$ while having comparable charging energies to InGaAs. Si4 displays the most prominent $N\!=\!4$ shell, best matching simulation, as well as having a distinct shell at $N\!=\!12$. This is likely due to improvements in our process leading to more symmetric metallization of the dot gates. The shell structure that we observe in InGaAs is the result of two-dimensional parabolic confinement and spin degeneracy,\cite{Tarucha,Reimann} whereas that for Si indicates an additional two-fold valley degeneracy as expected.

\begin{figure}
\begin{centering}
\includegraphics[width=3.2in]{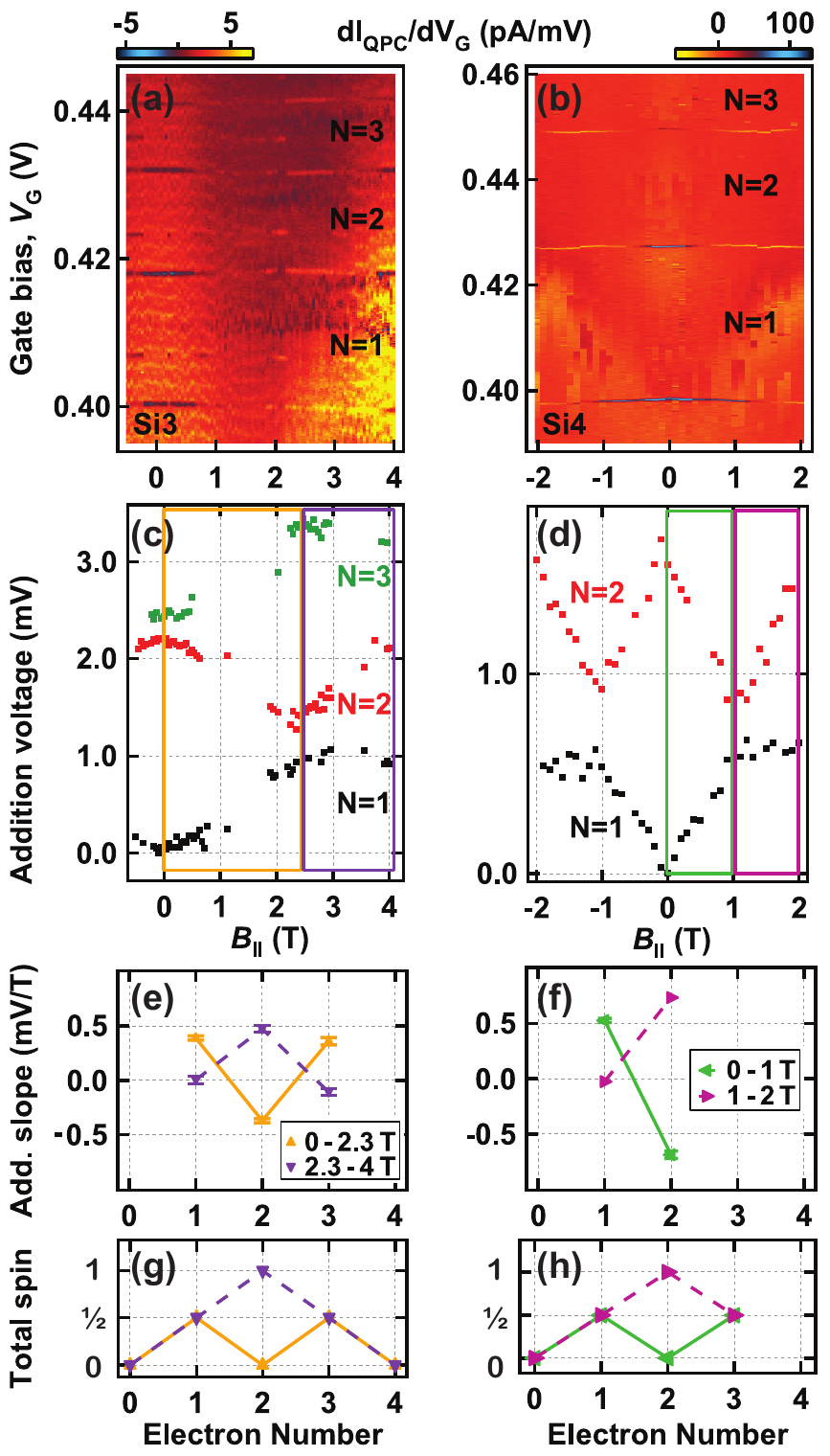}
\par\end{centering}
\caption{\label{fig:Fig3}[(a) and (b)] Plot of differential transconductance, $dI_\text{QPC}/dV_\text{G}$, vs gate voltage and in-plane magnetic field for devices Si3 and Si4. [(c) and (d)] Extracted addition voltages versus magnetic field, offset for clarity. [(e) and (f)] Slopes of linear fits to the boxed sections of data in (c) and (d) (indicated by matching colored lines.) [(g) and (h)] Total spin implied by (e) and (f).}
\end{figure}

The sequences of spin fillings for these devices are obtained by
performing magnetospectroscopy measurements. The application
of an in-plane magnetic field, $B_{\|}$, modifies the energetics
of the $N$-electron system primarily through the Zeeman term in the
Hamiltonian. Measuring the dots' transitions as a function of both
gate voltage and magnetic field allows inference of the total projected
spin state of the dot as a function of electron number for various
$B_{\|}$.\cite{Weis,Potok,Hanson} In Figs.~\ref{fig:Fig3}(a) and \ref{fig:Fig3}(b) we
plot the differential transconductance for devices
Si3 and Si4 versus gate bias and stepped magnetic field, collected 
in largely the same manner as for the addition
spectra of Figs.~\ref{fig:Fig2}(a) and \ref{fig:Fig2}(b).
The centers of each transition versus
$B_{\|}$ are extracted from the data by fitting the surrounding 2
mV data to a Gaussian function with fixed linewidth. 
All fits with transition
center uncertainties less than a fixed threshold are included
in subsequent analysis. In order to remove spin-independent
effects on the chemical potentials,\cite{Weis} the separations between
transitions are calculated and plotted in Figs.~\ref{fig:Fig3}(c) and \ref{fig:Fig3}(d).
Slopes from piecewise linear fits for magnetic fields in the four boxed regions of Figs.~\ref{fig:Fig3}(c) and \ref{fig:Fig3}(d), are shown along with error bars in Figs.~\ref{fig:Fig3}(e) and \ref{fig:Fig3}(f), consistent with multiples of the expected $g\mu_B/\alpha_\text{Si3(Si4)}\approx0.52\pm0.05(0.67\pm0.07)~\text{mV/T}$ for $g\!=\!2$.
Taking $S(0)\!=\!0$ and $S(1)\!=\!\nicefrac{1}{2}$, the total spin state can be inferred from the addition slopes, Figs.~\ref{fig:Fig3}(g) and \ref{fig:Fig3}(h).\cite{Weis,Potok,Hanson}

The $N\!=2\!$ spin state of device Si3 demonstrates a singlet-triplet (S-T) transition at $B_{\|}=2.3\pm0.1~\text{T}$ as evidenced by the change from the spin filling sequence $0\rightarrow\nicefrac{1}{2}\rightarrow0$ at low fields to $0\rightarrow\nicefrac{1}{2}\rightarrow1$ at higher fields. Device Si4 demonstrates an identical S-T transition at a lower magnetic field $B_{\|}=1.0\pm0.1~\text{T}$. For devices Si1 and Si2 only a triplet state was observed down to zero magnetic field.\cite{Si1Si2} Device InGaAs2 displayed the expected Hund's rule spin filling out to the third shell.\cite{InGaAs2,Tarucha}

The magnetic field corresponding to the S-T transition indicates the energy of the lowest lying triplet state; from $\Delta E_\text{S-T}=g\mu_{B}B_{\|}$ we find $\Delta E_\text{S-T}=270$ and $120~\mu\text{eV}$ for device Si3 and Si4, respectively. As the intervalley exchange energy is negligible, this is a strict lower bound on the valley splitting, $\Delta E_\text{valley}\geq\Delta E_\text{S-T}$.\cite{Hada} The equality likely holds due to the large orbital energies of these high-symmetry quantum dots. 
Further validation is provided by our calculations of the valley splitting in a microscopically-based framework which directly incorporates into the effective mass calculation the core physical origin of intervalley mixing --- the change to the local crystal potential due to substitution of individual host atoms in a heterostructure. A value $\Delta E_\text{valley}\approx 250~\mu\text{eV}$ is predicted for the ground orbital in both Si3 and Si4 devices for the case of perfectly flat and atomically sharp heterointerfaces; the splitting would be reduced in the presence of interface imperfections.

We have demonstrated high-symmetry quantum dots in Si/SiGe and InGaAs, with occupation number controllable from $N\!=\!0$. Magnetospectroscopy measurements reveal a valley non-degenerate ground state for two Si dots, a critical requirement for the eventual control and manipulation of single spins in this system.\cite{Eriksson,Hanson}

The authors gratefully acknowledge Thaddeus Ladd and Professor Charles Marcus for many useful discussions. Sponsored by United States Department of Defense. The views and conclusions contained in this document are those of the authors and should not be interpreted as representing the official policies, either expressly or implied, of the United States Department of Defense or the U.S. Government. 
Approved for public release, distribution unlimited.

\end{document}